\documentstyle[aps,epsfig,prl,multicol]{revtex}
\begin{document}
\vspace{0.0cm}
\draft

\title{Time dependence of occupation numbers and thermalization time
in closed chaotic many-body systems}

\author{V.V. Flambaum $^1$ \thanks{email address:
flambaum@newt.phys.unsw.edu.au} and F.M. Izrailev $^2$}

\address{$^1$ School of Physics, University of New South Wales,
Sydney 2052, Australia\\ $^2$Instituto de F\'{\i}sica, Universidad
Aut\'onoma de Puebla, Apartado Postal J-48, Puebla 72570,
M\'exico}

\date{\today}
\maketitle

\begin{abstract}

We study the time evolution of occupation numbers for interacting
Fermi-particles in the situation when exact compound states are
``chaotic". This situation is generic for highly excited
many-particles states in heavy nuclei, complex atoms, quantum
dots, spin systems and quantum computer models. Numerical data
show perfect agreement with a simple theory for the onset of
thermalization in close systems of interacting particles.

\end{abstract}


\begin{multicols}{2}

It is known that highly excited states can be treated as ``chaotic"
ones in many-body systems, such as complex atoms \cite{Ce},
multicharged ions \cite {ions}, nuclei \cite{nuclei}, spin
systems \cite{Nobel,spins} and quantum computer models
\cite{comp,F2000}. This happens due to a very high density of
many-particle states which strongly increases with an increase of
energy. For example, in the case of $n$ Fermi-particles occupying
the finite number $m$ of ``orbitals" (single-particle states), the
total number $N$ of many-body states grows exponentially fast with
an increase of number of particles, $N =m!/n!(m-n)!\sim \exp(c_0
n)$. Correspondingly, the density $\rho_f$ of those many-body
states which are directly coupled by a two-body interaction, also
grows very fast. Therefore, even a relatively weak interaction
between the particles can lead to a strong mixing between
unperturbed many-body states (``basis states"). As a result, an
exact (perturbed) eigenstate is represented by a chaotic
superposition of a large number of components of basis states
\cite{alt,FI00,hans}.
The number of principal basis components in such {\it chaotic
eigenstates} can be estimated as $N_{pc}\sim \Gamma /D$ where $\Gamma
$ is the {\it spreading width} of a typical component that can be
expressed through the Fermi golden rule, and $D$  is the
mean interval between the many-body levels.

In this paper we extend the quantum
chaos approach to the problem of a time evolution of an initially
excited basis state. This initial state may contain 
 one or several excited electrons above other electrons
in the ground state of  a quantum dot or atom.

Exact many-body eigenstates $\left| k\right\rangle \,$ of a 
Hamiltonian $H=H_0+V$ of interacting Fermi-particles can be
expressed in terms of simple {\it shell-model basis states}
$\left| f\right\rangle $ of $ H_0$,
\begin{equation}
\label{slat}\left| k\right\rangle =\sum\limits_f C_f^{(k)}\left|
f\right\rangle \,;\,\,\,\,\,\,\,\,\left| f\right\rangle
=a_{f_1}^{+}...a_{f_n}^{+}\left| 0\right\rangle.
\end{equation}
Here $\left|0\right\rangle$ is the ground state, $a_s^{+}$ is the
creation operator and $C_f^{(k)}$ are components of an exact
chaotic eigenstate $\left| k\right\rangle$ in the unperturbed basis, 
that is formed by a residual interaction $V$.

Below we consider the time evolution of the system, assuming that
initially ($t=0$) the system is in a specific basis state $\left|
i\right\rangle $. This state can be expressed as a sum over exact
eigenstates,
\begin{equation}
\label{in}\left| i\right\rangle =\sum\limits_kC_i^{(k)}\left|
k\right\rangle,
\end{equation}
therefore, the time-dependent wave function reads as
\begin{equation}
\label{psit}\Psi (t) =\sum\limits_{k,f}C_i^{(k)}C_f^{(k)}\left|
f\right\rangle \exp(-i E^{(k)}t).
\end{equation}
Here $E^{(k)}$ are the eigenvalues corresponding to the
eigenstates $|k>$. The sum is taken over the eigenstates $|k>$ and
basis states $|f>$ (hereafter we assume $\hbar=1$).

The occupation number $n_\alpha$ of a single-particles state
$\alpha$ for the wave function (\ref{psit}) of the system is
given by the expression,
\begin{equation}
\label{nt}
n_\alpha=\left\langle \Psi (t) | \hat{n}_\alpha| \Psi
(t)\right\rangle =
\sum_q \left\langle q | \hat{n}_\alpha| q \right\rangle
\left ( S_q^{(d)} + S_q^{(fl)} \right ).
\end{equation}
with $\hat{n}_\alpha=a_\alpha^\dag a_\alpha $. Here $S_q^{(d)}$ is the
diagonal term,
\begin{equation}
\label{diag}
S_q^{(d)}= \sum_k |C_i^{(k)}|^2 |C_q^{(k)}|^2,
\end{equation}
and $S_q^{(fl)}$ stands for off-diagonal terms,
\begin{equation}
\label{offd}
S_q^{(fl)}= \sum_{k\neq p} C_i^{(k)}C_q^{(k)}C_i^{(p)}C_q^{(p)}
\exp{\left [ i\left (E^{(k)}-E^{(p)}\right)t \right ]}.
\end{equation}

Since at $t=0$ the only unperturbed basis state $|i\rangle$ is
excited, the occupation numbers $n_\alpha(t)$ are equal to $0$ or
$1$ (for Fermi systems). For large times, $t \rightarrow \infty$,
the terms in the  off-diagonal sum $S_q^{(fl)}$ oscillate very rapidly as a
function of spacings $E^{(k)}-E^ {(p)}$, therefore,
$S_q^{(fl)}$ tends to zero. Thus, the asymptotic distribution
$n_\alpha (\infty)$ of occupation numbers is determined by the
diagonal term only.

If the number of principal components in the wave function is
large, one can replace $S_q^{(d)}$ by its average value,
\begin{eqnarray}
\label{Smean}
\overline{S_q^{(d)}} = \widetilde {F}(E_i , E_q) \approx
\sum_k F(E_i , E^{(k)}) F(E_q , E^{(k)})\\
 \approx \int F(E_i , E) F(E_q , E) \rho(E) dE.
\end{eqnarray}
Here $\rho(E)=D^{-1}(E)$ is the density of energy spectrum, and
\begin{equation}
\label{Ffun}
F(E_i , E^{(k)}) \equiv \overline{| C_i^{(k)}|^2}
\end{equation}
is the so-called {\it $F$-function} that characterizes
the shape of eigenstates in the unperturbed basis. This
$F$-function has been studied in details for different models
(see, e. g. \cite{BM69,Ce,nuclei,FI97,Kota} and the review \cite{review}).
 As was shown in \cite{FI00,casa}, in many cases the $F$-function can be
approximated by the following expression (see also \cite{decay}),
\begin{equation}
\label{FE}
F(E_i,E) \rho(E) = B \frac{\exp {\left
[-\frac{(E-E_c)^2}{2\sigma^2} \right ]}} {(E-E_i)^2
+\Gamma_i^2/4}.
\end{equation}
In this expression $B$ is the normalizing constant that can be
found from the relation $\int F(E_i,E) \rho(E) dE=1$, and
$\Gamma_i$ is the {\it spreading width} of a compound state
$|i\rangle$. The energy $E_c$ stands for the
center of effective energy band $\sigma$ of the Hamiltonian matrix, 
and may not coincide with $E_i$.

If residual interaction $V$ is not very strong,
the spreading width is determined by the standard golden rule,
$\Gamma_i \approx 2 \pi
\overline{|V_{if}|^2}\rho_f\ll \sigma $. 
Here $\rho_f$ is 
the density of final basis states $|f\rangle$ that are
coupled to $|i\rangle$ by the interaction $V$.
In the limit $\Gamma \ll \sigma$ the parameter $\sigma^2$
is equal to the variance
$\sigma_f^2$ of  $\rho_f$.
In the opposite limit $\Gamma \gg \sigma$
the value of $\sigma^2$ is approximately 
the same as the variance $\Delta_E^2$ of
the strength function
$F(E_i,E) \rho(E)$,  see Ref.\cite{FI00}. Note also, that $\sigma^2$
is always finite due to a finite range of an interaction in
the energy representation.

The form of
$F(E_i,E)$ in this limit $\Gamma \ll \sigma$ has simple
Breit-Wigner (BW) shape (apart from long tails that are highly
non-universal \cite{Ce}),
\begin{equation}
\label{BW}
F(E_i,E)= \frac{1}{2\pi \rho} \frac{\Gamma}{(E-E_i)^2 +
\Gamma^2/4},
\end{equation}
where $\Gamma \equiv \Gamma_i$.
In this case the function $\widetilde{F}(E_i,E_q)$ that gives the
final shape of the time-dependent wave function, has also the BW
form (\ref{BW}) with the width $\widetilde{\Gamma}=2\Gamma$. Note
that fluctuations of the diagonal term $S_f^{(d)}$ are small,
$S_f^{(d)}-\overline{S_f^{(d)}} \sim
\overline{S_f^{(d)}}/{\sqrt {N_{pc}}}$. Thus, for a large number
of principal components $N_{pc}$ the occupation numbers both in
compound eigenstates and in large-time asymptotic of
the wave function, have the
self-average property. As a result, we obtain the following
expression for the asymptotic values of occupation numbers,
\begin{equation}
\label{FT}
n_\alpha(\infty) = \sum_q \langle q|\hat{n}_\alpha|q\rangle
\widetilde {F}(E_i,E_q).
\end{equation}
In fact, this distribution is very close to the distribution of
occupation numbers in compound eigenstates of a system,
\begin{equation}
\label{FD}
\hat{n}_\alpha = \sum_q \langle q|\hat{n}_\alpha|q\rangle
F(E_q,E).
\end{equation}
The properties of the distribution (\ref{FD}) have been studied 
in details \cite{Ce,nuclei,FIC96,FI97,ions,ponomarev} for different
models.

In many cases (the two-body random interaction (TBRI)
model \cite{FIC96,FI97}, $s-d$
nuclear shell model \cite{nuclei}, multiply charged ions \cite{ions})
the $F$-distribution (\ref{FD}) is very close to the standard
Fermi-Dirac distribution for the occupation numbers with a certain
temperature $T(E)$. In these cases the asymptotic distribution
(\ref{FT}) of the occupation numbers in the time-dependent problem
is also given by the Fermi-Dirac distribution. In other cases,
like $Ce$ atom \cite{ponomarev} where the residual interaction between
particles on some orbitals is larger than the distance between
single-particle energy levels, the $F$-distribution (\ref{FD})
deviates very strongly from the Fermi-Dirac distribution. This
happens since the mean-field approximation is not a ``good" one.
However, even in this case the distribution of $n_\alpha$ is close
to that obtained from the canonical distribution that takes into
account a repulsion between electrons (see details in \cite{ponomarev}).
This canonical distribution is also characterized by some 
temperature $T(E)$. Thus, it is natural to term the time evolution
of $n_\alpha(t)$ as the process of {\it thermalization} in a closed
system.

The natural question is: how fast is this termalization? To answer
this question, one needs to study the evolution of a many-body
system taking into account a two-body nature of the interaction
between particles. For this, it is convenient to consider the
so-called {\it cascade
model} (see details in \cite{entrop}), that is based on the
representation of unperturbed many-body states in the form of ``classes". 
In this picture, the first class contains those
$N_1$ basis states that are directly coupled by the two-body
interaction $V_{if}$. Correspondingly, the second class consists
of $N_2$ basis states that are coupled with the initial state in
the second order of the interaction $V$ (therefore, the coupling
is proportional to $V_{ik}V_{kf}$), etc. 

The analytical expression for time dependence
$W_n(t)$ of a population of the class $n$ has been obtained in
Ref.\cite{entrop}. The population in the ``zero" class is just a
probability $W_0(t)$ of the system to remain in an initial state.
In the BW case, $\Gamma \ll \sigma$, this probability 
for large times is
$W_0(t)\sim\exp(-\Gamma t)$. In other limiting case, $\Gamma
\gtrsim \sigma$, when the form of the $F$-function 
is close to the Gaussian, the time dependence $W_0(t)$ is more
complicated \cite{decay}. For example, for  $\Gamma \gg \sigma$, it
was found to be $W_0(t)=\exp(-\Delta_E^2 t^2)$, up to a very  long time.
In general case it changes from $W_0(t)=\exp(- \Delta_E^2 t^2)$
for small times,  to $W_0(t)=C\exp(-\Gamma t)$ for large times
and saturates near $W_0(t) \approx 3/N_{pc}$ (here 
$C>1$ is the time-independent
coefficient, see details in \cite{decay}).  

Let us now consider the thermalization process. Initially, the
occupation numbers $n_\alpha (0)$ are equal to $0$ or $1$. In the
BW regime, the first class is populated during the time
$\tau \sim 1/\Gamma$ \cite{entrop} (or, $\tau \sim 1/\Delta_E$
in the Gaussian regime, for 
$\Gamma \gtrsim \sigma$). At that time  the occupation numbers
$n_\alpha$ already strongly deviate from their initial values
since the two-body interaction can move any of two particles to
new single-particles levels characterized by the
energies $\epsilon_\alpha$. Thus, the characteristic time for a  
thermalization of the occupation numbers is determined by the population
 time $\tau$ 
for the first class. Note, that this time is 
also a characteristic time for the
``decay" of the probability $W_0(t)$.
To compare with, the population of all $n_c$ classes in a particular
system requires a longer time $\tau_{n_c} \sim n_c \tau$
\cite{entrop}. Thus, in the case of $n_c \gg 1$ (e.g. in a mesoscopic system)
the thermalization of the occupation numbers may occur on a  smaller time
scale than the onset of a complete statistical equilibrium.

This suggests a simple derivation of the time dependence
$n_\alpha(t)$ for occupation numbers. From the normalization
condition $\sum_{s=0}^{n_c} W_s =1$ one can find the population of
all classes with $s\neq 0$ as $\sum_{s=1}^{n_c} W_s = 1-W_0$.
Now we assume that the thermalization of the occupation numbers
occurs on the time scale $\tau$. This assumption leads
to a simple expression
\begin{equation}
\label{appr}
n_\alpha(t) = n_\alpha(0) W_0(t) + n_\alpha (\infty) \left (
1-W_0(t) \right).
\end{equation}
Here $n_\alpha(0)$ are initial occupation numbers ($0$ or $1$),
and $n_\alpha(\infty)$ determines equilibrium occupation numbers
(e.g., the Fermi-Dirac distribution).

In Figs.1-2 we compare numerical data for $n_\alpha(t)$ with the
above estimate (\ref{appr}). Two situations are studied, the first
(Fig.1) corresponds to the BW regime (weak interaction regime,
$\Gamma \ll \sigma$) , and in the second (Fig.2) the form of the 
$F$-function is
close to the Gaussian ($\Gamma \gtrsim \sigma$). For numerical
simulation, we have used the model with random two-body
interaction (see details, in \cite{FI97}) with $n=6$
Fermi-particles that occupy $m=12$ single-particle levels. Since
the analytical expression for $W_0(t)$ in general case is quite
complicated, in numerical simulations we use the exact value of
$W_0(t)$ computed numerically.

Overall, there is an excellent agreement between the theory and
numerical data (apart from fluctuations that are neglected in the
theory). In order to simplify the picture, we have taken the
initial basis state $|i\rangle$ just in the middle of the
many-particle spectrum that consists of $924$ levels. In this case
all final values of the occupation numbers are equal,
$n_\alpha(\infty)=n/m=1/2$ that corresponds to the infinite
temperature $T$.

\begin{figure}[htb]
\vspace{-1.0cm}
\begin{center}
\hspace{-1.5cm}
\epsfig{file=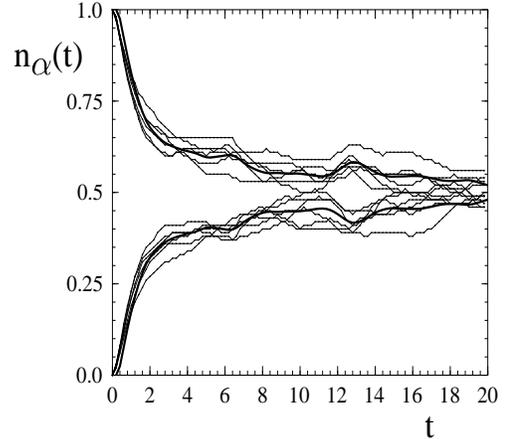,width=3.0in,height=2.3in,angle=-90}
\vspace{-0.8cm}
\narrowtext
\caption{
Time dependence of occupation numbers $n_\alpha(t)$ for the BW
regime (weak interaction ).
Thin curves with dots present numerical data, thick smooth
curves correspond to the analytical expression (\ref{appr}), see
the text. Computations are made for the model with random two-body
interaction, with $n=6, m=12, \eta \equiv <V^2/D_0^2> \approx
0.003, \Gamma \approx 0.50$, $\Delta_E \approx 1.16$ (here $D_0$ is
the mean spacing between single-particle energies).
}
\end{center}
\end{figure}

\begin{figure}[htb]
\vspace{-1.0cm}
\begin{center}
\hspace{-1.5cm}
\epsfig{file=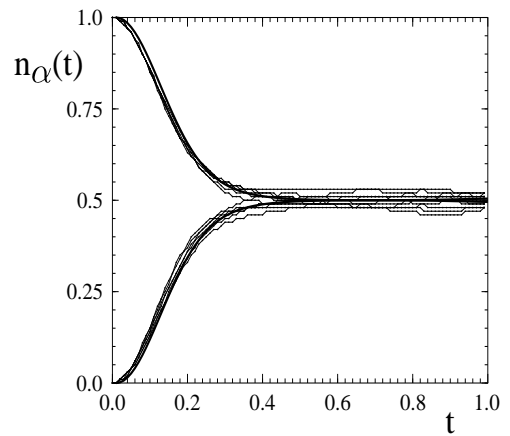,width=3.0in,height=2.3in,angle=-90}
\vspace{-0.8cm}
\caption{
Time dependence of occupation numbers $n_\alpha(t)$
for the case when the strength function has the gaussian form:
$\eta \equiv
<V^2/D_0^2>
\approx 0.083, \Gamma \approx 10.5, \Delta_E \approx 5.8$.
}
\end{center}
\end{figure}

We would like to point out on an interesting difference between
two cases of a weak and strong interaction between the particles.
Specifically, for a weak interaction (Fig.1) the transition to
equilibrium values of $n_\alpha$ has a character of damped
oscillations. The number of principal components $N_{pc} \sim
\Gamma \rho^{-1}$ in this case is not very large, this is why there are
considerable fluctuations in $n_\alpha(t)$ even in the
equilibrium. To compare with, the case of a strong interaction
(Fig.2) shows fast and monotonic transition to thermal values of
$n_\alpha(\infty)$ with relatively small fluctuations, see also \cite{entrop}.

  Our results demonstrate that actually there are two time scales
in the onset of thermalization. The first one is determined by $\tau$.
It characterizes an
``initial thermalization'', and  allows one to use Eq.(\ref{appr})
for a description of the time dependence of occupation numbers. 
For larger times, 
damped quantum oscillations (with period $T\sim n_c \tau$) may
occur in the transition to a complete
equilibrium \cite{decay,entrop}.

In conclusion, we present a simple  theory for the onset
of thermalization in closed systems of interacting particles. The
theory allows to describe the time dependence of the occupation
numbers, and shows the relaxation to an equilibrium distribution.
Numerical data obtained for the model of a random two-body
interaction between finite number of Fermi-particles, demonstrate
a very good correspondence to the theoretical predictions. Our
results can be used in different applications, such as complex
atoms, heavy nuclei, atomic clusters, quantum dots, 
etc.

 This work was supported by the Australian
Research Council. One of us (F.M.I.) gratefully acknowledges the
support by CONACyT (Mexico) Grant No. 34668-E.

\end{multicols}

\begin{thebibliography}{99}
\bibitem{Ce}  V. V. Flambaum, A. A. Gribakina, G. F. Gribakin, and M. G.
Kozlov, Phys. Rev. {\bf A 50}, 267 (1994).

\bibitem{ions}  G.F. Gribakin, A.A. Gribakina, V.V. Flambaum.
Aust.J.Phys.{\bf 52}, 443 (1999).

\bibitem{nuclei}  M.Horoi, V.Zelevinsky and B.A.Brown,
Phys. Rev. Lett. {\bf 74}, 5194 (1995); V.Zelevinsky, M.Horoi and
B.A.Brown, Phys. Lett. {\bf B} {\bf 350}, 141 (1995); N.Frazier,
B.A.Brown and V.Zelevinsky, Phys. Rev. {\bf C} {\bf 54}, 1665
(1996); V.Zelevinsky, B.A.Brown, M. Horoi and N.Frazier, Phys.
Rep., {\bf 276} , 85 (1996).

\bibitem{Nobel}  V. V. Flambaum, Proc. 85th Nobel Symposium, Phys. Scr.
{\bf 46}, 198 (1993).

\bibitem{spins}  B.~Georgeot and D.~L.~Shepelyansky, Phys. Rev. Lett.
{\bf 81}, 5129 (1998).

\bibitem{comp} B. Georgeot and D.L. Shepelyansky, Phys. Rev. E.,
{\bf 62}, 3504 (2000); ibid, 6366; G.P.Berman, F.Borgonovi,
F.M.Izrailev, and V.I.Tsifrinovich, quant-ph/0012106; ibid,
0104086; P.G. Silvestrov, H. Schomerus, and C.W.J. Beenakker,
quant-ph/0012119.

\bibitem{F2000}  V.V. Flambaum. Aust.J.Phys. {\bf 53}, N4, (2000);
quant-ph/9911061.

\bibitem{alt}  S. Aberg. Phys. Rev. Lett. {\bf 64}, 3119 (1990).
D.L.Shepelyansky and O.P.Sushkov, Europhys. Lett. {\bf 37}, 121
(1997); B.L.Altshuler, Y.Gefen, A.Kamenev and L.S.Levitov, Phys.
Rev. Lett., {\bf 78}, 2803 (1997); A.D.Mirlin and Y.V.Fyodorov,
Phys. Rev. {\bf B 56}, 13393 (1997); D.Weinmann, J.-L. Pichard and
Y.Imry, J.Phys. I France, {\bf 7}, 1559 (1997); P.Jacquod and
D.L.Shepelyansky, Phys. Rev. Lett. {\bf 79} , 1837 (1997);
V.V.Flambaum and G.F.Gribakin, Phys. Rev. {\bf C 50}, 3122 (1994);
P.G.Silvestrov, Phys. Rev. Lett., {\bf 79}, 3994 (1997); Phys.
Rev. E. {\bf 58}, 5629 (1998).

\bibitem{FI00}  V.V.Flambaum and F.M.Izrailev, Phys. Rev. {\bf E 61},
2539 (2000).

\bibitem{hans} T.Rupp, H.A.Weidenm\"uller and J.Richert, nucl-th/0003053;
L.Benet and H.A.Weidenm\"uller, cond-mat/0005103; L.Benet, T.Rupp and
H.A.Weidenm\"uller, cond-mat/0010425.

\bibitem{BM69}  A. Bohr and B. Mottelson, {\em Nuclear structure, Vol. 1}
(Benjamin, New York, 1969).

\bibitem{FI97}  V.V.Flambaum and F.M.Izrailev, Phys. Rev. {\bf E 55},
R13 (1997); {\bf E 56}, 5144 (1997).

\bibitem{Kota}  V.K.B. Kota and R. Sahu, nucl-th/0006079.

\bibitem{review}
V.K.B. Kota, Phys.Rep. {\bf 347}, 223 (2001).

\bibitem{casa}  G.Casati, V.V.Flambaum, and F.M.Izrailev, (2000), unpublished.

\bibitem{decay} V.V. Flambaum and F.M. Izrailev,
Phys. Rev. {\bf E 64}, 026124 (2001).

\bibitem{ponomarev} V.V. Flambaum, A.A. Gribakina, G.F. Gribakin,
 I.V. Ponomarev,  Phys. Rev. {\bf E 57},4933 (1998);
 Physica {\bf D 131}, 205 (1999).

\bibitem{FIC96} V. V. Flambaum, F. M. Izrailev,
and G. Casati,  Phys. Rev. {\bf E 54}, 2136 (1996).

\bibitem{entrop}  V.V.Flambaum and F.M.Izrailev, quant-ph/0103129,
 Phys. Rev. E, in press.

\end{thebibliography}
\end{document}